\documentclass[aps,prl,superscriptaddress,amsfonts,amsmath,amssymb,showpacs,floatfix,reprint,longbibliography,footinbib]{revtex4-2}

\usepackage{url}
\usepackage{bm,bbm}
\usepackage{graphicx}
\usepackage{color}
\usepackage{xcolor}
\usepackage[colorlinks=true, urlcolor=blue, linkcolor=blue, citecolor=blue, pdftex]{hyperref}
\usepackage[english]{babel}
\usepackage{physics}
\usepackage{slashed}

\usepackage{feynmp-auto}

\newcommand{\betaeff}{\beta_{\rm eff}}

\usepackage{soul}
\usepackage{blindtext}
\usepackage{lipsum}
\usepackage{nicefrac}

\newcommand{\todo}[1]{}

\begin{document}

\title{Persistence of the Berezinskii--Kosterlitz--Thouless transition\\with long-range couplings}

\author{Luis Walther}
\thanks{lwalther@pks.mpg.de}
\affiliation{Technical University of Munich, TUM School of Natural Sciences, Physics Department, 85748 Garching, Germany}
\affiliation{Max-Planck-Institut f\"{u}r Physik komplexer Systeme, 01187 Dresden, Germany}
\author{Josef Willsher}
\affiliation{Technical University of Munich, TUM School of Natural Sciences, Physics Department, 85748 Garching, Germany}
\affiliation{Munich Center for Quantum Science and Technology (MCQST), Schellingstr. 4, 80799 München, Germany}
\affiliation{Max-Planck-Institut f\"{u}r Physik komplexer Systeme, 01187 Dresden, Germany}

\author{Johannes Knolle}
\affiliation{Technical University of Munich, TUM School of Natural Sciences, Physics Department, 85748 Garching, Germany}
\affiliation{Munich Center for Quantum Science and Technology (MCQST), Schellingstr. 4, 80799 München, Germany}
\affiliation{Blackett Laboratory, Imperial College London, London SW7 2AZ, United Kingdom}

\date{\today}

\begin{abstract}
The Berezinskii--Kosterlitz--Thouless (BKT) transition is an archetypal example of a topological phase transition, which is driven by the proliferation of vortices. In this Letter, we analyze the persistence of the BKT transition in the XY model under the influence of long-range algebraically decaying interactions of the form $\sim 1/{r^{2+\sigma}}$.
The model hosts a magnetized low temperature phase for sufficiently small $\sigma$. Crucially, in the presence of long-range interactions, spin waves renormalize the interaction between vortices, which stabilizes the BKT transition. As a result, we find that there is no direct transition from the magnetized to the disordered phase and that the BKT transition persists for arbitrary long-range exponents, which is distinct from previous results. 
We use both Landau--Peierls-type arguments and renormalization group calculations---including a coupling between spin wave and topological excitations---and obtain similar results. We emphasize that Landau--Peierls-type arguments are a powerful tool for analyzing continuous spin models. 
We discuss the relevance of our findings for current Rydberg atom experiments, and highlight the importance of long-range couplings for other types of topological defects.
\end{abstract}

\maketitle 
\paragraph{Introduction.}\label{sec:intro}
One of the key aims of condensed matter physics is to chart out phase diagrams and understand phase transitions. The two-dimensional XY spin model holds a special place in this quest: It hosts a phase transition driven by topological defects.
This transition was first discovered by Berezinskii \cite{Berezinsky:1970fr} and Kosterlitz and Thouless \cite{Kosterlitz_1973nobelpaper}, and is therefore termed the Berezinskii--Kosterlitz--Thouless (BKT) transition. It is a surprising result, since the Mermin--Wagner theorem forbids phase transition described by Landau's picture for systems with continuous symmetries in two dimensions \cite{MerminWagnerTheorem}. The BKT transition is a topological phase transition and the first known example of a transition outside the scope of Landau's picture. 
Although first theoretically observed in spin systems, the BKT mechanism has been found to describe many other systems; its physics appears in Josephson Junction arrays \cite{Capriotti2004}, in thin $^4\text{He}$ films \cite{PhysRevLett.40.1727}, in thin film superfluids \cite{doi:10.1126/science.abq6753}, in two-dimensional Coulomb gases \cite{PhysRevLett.46.1006, doi:10.1126/science.1146006, PhysRevB.94.085104} and in classical two-dimensional $Z_n$ clock models for $n>4$ \cite{PhysRevE.101.062111, Goswami_2025}.
%%%%%%%%%%%%%%%%%%%%%%%%%%%%%%%%%%%%%%%%%%%%%%%%%%%%%%%%%%%%%%%%%%%%%%%%%%%%%%%%%%%%%%%%
\begin{figure*}[!t]
    \centering
    \includegraphics[width=1\textwidth]{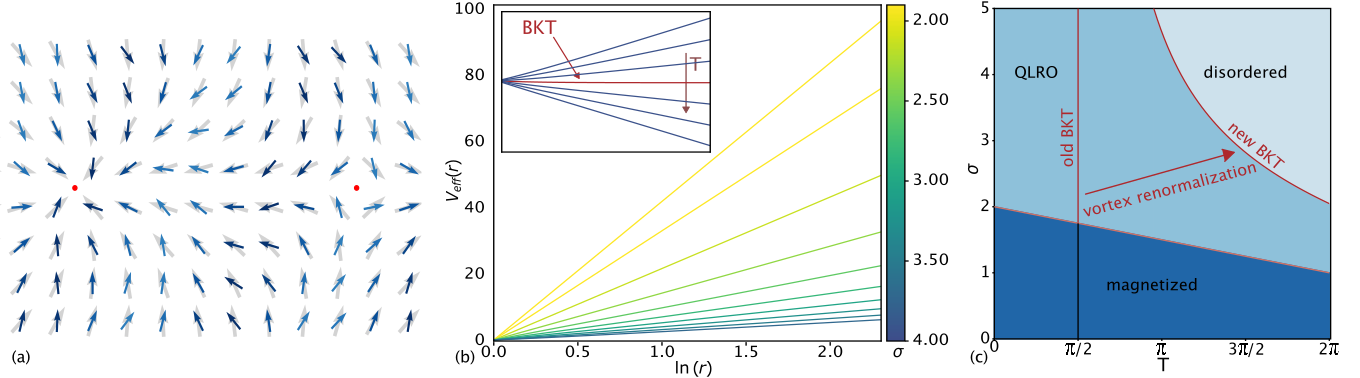}
    \caption{(a) Vortex-antivortex configuration (grey) overlayed with a spin wave configuration (blue). The vortices are separated at a distance of $r$ in units of the lattice spacing $a$. In the short-range XY model, spin waves and vortices do not interact with each other, which is no longer true in the presence of long-range interactions. (b) Effective potential of a vortex-antivortex pair. Increasing the temperature decreases the incline of the slope until vortices deconfine at the BKT transition (see inset figure, incline becomes zero). Long-range interactions strengthen the vortex-antivortex potential by increasing the slope. (c) Phase diagram of the long-range XY model, as obtained by the renormalization of vortices by long-range interactions. The long-range interactions stabilize vortices, so that the BKT transition appears at higher temperatures. As a result, the QLRO phase persists at intermediate temperatures for arbitrarily low values of $\sigma$.}
    \label{fig:figure1}
\end{figure*}
%%%%%%%%%%%%%%%%%%%%%%%%%%%%%%%%%%%%%%%%%%%%%%%%%%%%%%%%%%%%%%%%%%%%%%%%%%%%%%%%%%%%%%%%

There has been a large recent interest in the study of long-range interacting systems, partly driven by recent advances in quantum simulator platforms~\cite{PhysRevLett.108.215301, PhysRevLett.108.210401, Lewis_2023,Yan2013,RevModPhys.82.2313, Schauß2012, Chen2023rydberg}. 
The influence of long-range interactions on spin systems has been studied for a long time in statistical and condensed matter physics \cite{Bouchet_2010, Defenu_2023}. In the presence of long-range interactions, the Mermin--Wagner theorem does not hold anymore, so that also in the XY model, a magnetized phase is possible \cite{Kunz1976}.
While the BKT mechanism appears naturally in many settings, it is not clear whether it is stable towards long-range interactions of the form $\sim 1/r^{d+\sigma}$.
In addition, long-range interactions can change the critical behavior of $d$-dimensional spin models, which has been extensively studied \cite{Kunz1976, fisherlongrangeexpo, sakcriterion, Defenu_2023, PhysRevE.89.062120, Brezin2014, defenucritinlongrangeonmodels, PhysRevE.92.052113}. In the case of long-range $\mathrm{O}(N)$ models, there exists a critical value $\sigma^*$ of the long-range exponent, such that the system displays short-range critical behavior for $\sigma>\sigma^*$, whereas long-range interactions are relevant for $\sigma<\sigma^*$. In a seminal contribution \cite{fisherlongrangeexpo}, Fisher et.al. first calculated $\sigma^*=2$. This value introduced a discontinuity in the critical exponents, which was later resolved by Sak in \cite{sakcriterion}: He proposed that $\sigma^*=2-\eta_{\rm SR}$, where $\eta_{\rm SR}$ is the spatial critical exponent of the short-range model. This is known as Sak's criterion. There has been considerable scrutiny around Sak's criterion, including many works agreeing with \cite{PhysRevE.89.062120, PhysRevLett.89.025703} or contradicting it \cite{PhysRevB.26.1336, Blanchard_2013}.

The case of the two-dimensional XY model is even more involved. In the short-range limit, there is no spontaneous symmetry breaking, and instead of a constant spatial critical exponent $\eta_{\rm SR}$, there is a whole critical phase, characterized by a temperature-dependent spatial critical exponent $\eta=1/(2\pi \beta)$, where $\beta$ is the inverse temperature.
Further, Sak's criterion is expected to be challenged by the topological nature of the transition in the XY model.
Finally, when analyzing the phase structure of the short-range XY model, one often relies on duality mappings and approximations, such as the Villain approximation \cite{villainapproximation}. However, these mappings are known to no longer be valid for the long-range XY model (as for example the long-range Villain model has a different universality class from the long-range XY model \cite{Giachetti2023}).
As a result, the question about the influence of long-range interactions on the XY model has not been definitively settled.

In this work, we analyze the persistence of the BKT transition in the classical long-range XY model. The Hamiltonian of the model at hand reads
\begin{equation}\label{eq:lrxy_hamiltonian_original}
    H=\sum\displaylimits_{\langle i, j\rangle }\big[1-\cos{\big(\phi_i-\phi_j\big)}\big]+g\!\!\sum\displaylimits_{|i-j|>a}\!\!\frac{1-\cos{\big(\phi_i-\phi_j\big)}}{|i-j|^{2+\sigma}} \,.
\end{equation}
Here, $i, j\in \mathbf{Z}^2$ describe the two-dimensional lattice, and $\phi_i$ describes the phase of the local spin at lattice site $i$. 
The physical value of the long-range coupling is $g=1$, but we introduce it as a variable already to foreshadow a perturbative calculation later on.
% We have already introduced the long-range coupling $g$, foreshadowing that we will treat the long-range interaction perturbatively. Nevertheless, physically, we would set $g=1$.

We begin by summarizing what is known about the model: It hosts a magnetized phase at low temperatures for $\sigma\leq 2$ \cite{Kunz1976}. More recently, this bound has been refined to $\sigma \leq 2-\eta$~ \cite{giachettidefenu2021}. Relying both on Sak's argument \cite{sakcriterion} and the recent self-consistent harmonic approximation \cite{Giachetti_scha_2021}, it is expected that the system displays short-range behavior if $\sigma \geq 2$, including a BKT transition.
Whether the BKT transition survives in the regime $\sigma<2$ is less clear. In Ref.~\cite{giachettidefenu2021}, Giachetti \textit{et al.} argue that the BKT transition persists only up to $\sigma=7/4$. This value corresponds to $\eta_{SR}=1/4$ in Sak's framework, which is exactly the value of $\eta$ at the BKT transition $\beta_{c}=2/\pi$. Contrary to this, Xiao, Yao, \textit{et al.} recently published two numerical studies \cite{Xiao_2025, yao2025nonclassicalregimetwodimensionallongrange} on the long-range XY model, in which the authors find that the BKT transition only exists for $\sigma\geq2$. Using Monte Carlo methods, they find a second-order phase transition from a disordered to a magnetized phase for $\sigma<2$.
We aim to address this apparent disagreement by further investigating the 
% persistence of the BKT transition, using field-theoretic calculations. 
phase diagram using complementary field-theoretic methods.
% 
% We apply both Landau--Peierls-type (LP) arguments as well as renormalization group (RG) calculations. This also serves to introduce the method of LP arguments to be used for spin models with continuous symmetries.
We develop a Landau--Peierls-type (LP) method which can successfully be applied to spin models with continuous symmetries.
This, in turn, motivates us to take another look at the perturbative renormalization group \cite{giachettidefenu2021, Chaikin_Lubensky_1995, Kosterlitz_1973nobelpaper} at higher order, by including a coupling between topological and non-topological excitations.

LP-type arguments provide great physical intuition of the corresponding model and its excitations and phase transitions. The method was famously used by Peierls  \cite{Peierls_1936} to analyze the proliferation of domain walls in the two-dimensional Ising model, which leads to a symmetry breaking phase transition. The method can be applied to models displaying long-range interactions, such as the one-dimensional Ising model \cite{thoulesslrisingoned}. LP arguments can also be used in continuous spin systems \cite{Kosterlitz_1973nobelpaper}, where now topological defects take the role of domain walls.
The method considers the interplay of energy and entropy of the fundamental excitations of the model. Here, we do this by deriving an effective partition function of the vortex excitations which proliferate at the BKT transition.
LP arguments are in some sense complementary to RG calculations: While it is often impossible to determine universal exponents from LP arguments, it is much easier to chart out phase diagrams and access nonuniversal propeties like the critical temperature.

\paragraph{Landau--Peierls-type argument.}\label{sec:LP}
We now develop an LP argument for the long-range XY model. To do so, we derive an effective partition function of a vortex-antivortex pair. 
% We proceed similarly to \cite{giachettidefenu2021} and treat the long-range interactions perturbatively. 
We begin by transforming the Hamiltonian  Eq.~\eqref{eq:lrxy_hamiltonian_original} into the following form \cite{Chaikin_Lubensky_1995}
\begin{equation}\label{eq:hamil_writtenas_srandlr}
    \begin{split}
        &H=\frac{1}{2}\!\int\! \dd[2]{\bm{x}}\big(\nabla \varphi(\bm{x})\big)^2\!\!\!-\!\pi\! \sum\displaylimits_{\bm{x}\neq  \bm{y}}n_{\bm{x}}n_{\bm{y}}G(\bm{x}\!-\!\bm{y})\!+\! \sum_{\bm{x}}n_{\bm{x}}^2E_c\\
        &+\!\frac{g}{a^{2-\sigma}}\!\!\int\!\!\frac{\dd[2]{\bm{x}}\dd[2]{\bm{y}}}{|\bm{x}\!-\!\bm{y}|^{2+\sigma}}\!\big[1\!\!-\!\cos{\!\big(\varphi(\bm{x})\!-\!\varphi(\bm{x})\!+\!\theta(\bm{x})\!-\!\theta(\bm{x})\big)}\big].
    \end{split}
\end{equation}
Here, $n_{\bm{x}}$ is the vortex charge density, $G(\bm{r})\approx \ln(|\bm{r}|/a)$ is the Green's function of a two-dimensional Coulomb gas, $a$ is the lattice spacing, and $E_c$ is the core energy of vortices. The original spin orientation $\phi(\bm{x})$ has been split into a spin-wave field and a vortex field as $\phi(\bm{x})=\varphi(\bm{x})+\theta(\bm{x})$. 
We can observe that the long-range interactions act to couple vortices and spin waves, whereas they are fully decoupled in the short-range model. We formulate an effective partition function, which integrates over all configurations containing a vortex at position $\bm{r}_1$ and an antivortex at position $\bm{r}_2$:
\begin{equation}\label{eq:part_func_definition}
    Z=\int \dd[2]\bm{r}_1\dd[2]\bm{r}_2\int D[\varphi]e^{-\beta E_v(\bm{r}_1, \bm{r}_2, \varphi)}.
\end{equation}
From Eq.~\eqref{eq:hamil_writtenas_srandlr} we deduce the energy of a vortex-antivortex pair. We insert this into Eq.~\eqref{eq:part_func_definition} and obtain to lowest order in $g$:
\begin{equation}
     Z= 2\pi y^2Z_{\mathrm{sw}}\frac{L^2}{a^2}\int \frac{\dd r}{a}\left(\frac{r}{a}\right)^{1-2\pi\beta}e^{-\beta g \,  I\left(\frac{r}{a}\right)}.
\end{equation}
Here, $r=|\bm{r}_1-\bm{r}_2|$, $L^2$ is the size of the system, and $Z_{\rm sw}$ is the partition function of only spin waves, which is independent of $r$. The fugacity of vortices is $y=e^{-\beta E_c }$. We are left to evaluate $I\left(r/a\right)$, which is the long-range energy of a vortex-antivortex configuration after integrating out spin waves. This integral can be approximately solved, as shown in the End Matter. Inserting the solution, we find the final version of the effective partition function: 
\begin{equation}\label{eq:lp_effectivepart_final}
     Z= 2\pi y^2Z_{\rm sw}\frac{L^2}{a^2}\int \frac{\dd r}{a}\left(\frac{r}{a}\right)^{1-2\pi\betaeff}e^{-\beta g \, J\left(\frac{r}{a}\right)\cdot \left(\frac{r}{a}\right)^{2-\sigma-\eta}}.
\end{equation}
Here, $J$ only has a weak logarithmic dependence on its argument $r/a$, which can be neglected compared to the polynomial growth it is multiplied with. We define the renormalized $\betaeff$ as:
\begin{equation}
    \betaeff =\beta\left(1-\frac{\pi g}{2-\sigma-\eta}\right).
\end{equation}
%%%%%%%%%%%%%%%%%%%%%%%%%%%%%%%%%%%%%%%%%%%%%%%%%%%%%%%%%%%%%%%%%%%%%%%%%%%%%%%%%%%%%%%%
\begin{figure}[t!]
    \centering
    \includegraphics[width=\linewidth]{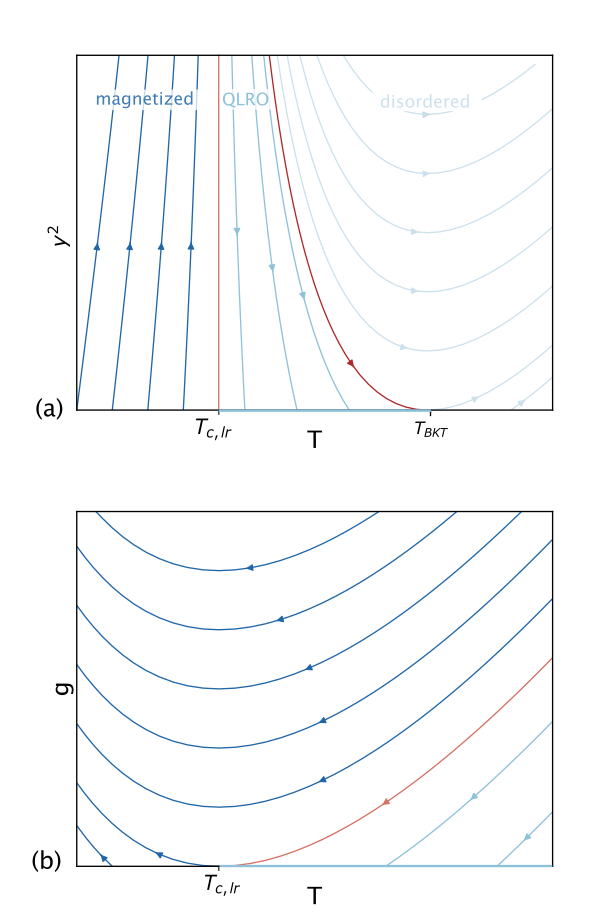}
    \caption{RG flow sections of the long-range XY model for $\sigma=1$. Note, the colors are chosen in accordance to the phase diagram in Figure \ref{fig:figure1}. (a) RG flow in the $T-y^2$-plane for fixed $g=1$. Because $g$ is kept fixed, also the effect of $g$ on the renormalization of $\beta$ is neglected. As can be seen, the BKT transition line persists up to the renormalized temperature $T_{{\mathrm{BKT}}}$ (see end matter for definition). At $T_{\mathrm{c, lr}}=2\pi(2-\sigma)$ long-range interactions become relevant, and from Eq.~\eqref{eq:rgflow_rgequations} we observe a divergence in the flow of $y^2$. We cannot make decisive statements what happens at this point, because both analytical and numerical treatments struggle with the divergence. The full analysis of the RG equations remains left open for future work.  (b) RG flow in the $T-g$-plane for $y^2=0$. As can be seen, long-range interactions become relevant for $T<T_{\mathrm{c, lr}}$.}
    \label{fig:rgflow}
\end{figure}
%%%%%%%%%%%%%%%%%%%%%%%%%%%%%%%%%%%%%%%%%%%%%%%%%%%%%%%%%%%%%%%%%%%%%%%%%%%%%%%%%%%%%%%%
We can now distinguish two cases: First, if $\sigma<2-\eta$, the exponential in Eq.~\eqref{eq:lp_effectivepart_final} suppresses configurations with a large separation of the topological defects. Here, long-range interactions are relevant which corresponds to a magnetized phase~\cite{giachettidefenu2021}. Second, if $\sigma>2-\eta$, the exponential converges to one for a large separation of the topological defects. Then, the behavior of vortices is determined by the exponent of the algebraic term. The behavior depends on $\beta$. We define the critical temperature $\beta_c$ of the BKT transition as
\begin{equation}
    \frac{2}{\pi}=\beta_{\rm{eff}, c}=\beta_c\left(1-\frac{\pi g}{2-\sigma-\eta_c}\right).
\end{equation}
For $\beta>\beta_c$, vortices are confined by the Coulomb interactions, and the system is in a quasi-long-range ordered phase. For $\beta<\beta_c$, vortices deconfine and the system is in a disordered phase.
The resulting phase diagram is shown in Fig.~\ref{fig:figure1}. Our key finding is that that $\beta_c$ increases for lower values of $\sigma$, thus the BKT transition does not interfere with the magnetization phase transition in the system. Consequently, the BKT transition persists for arbitrary values of the long-range exponent $\sigma$, and there is no direct transition from the magnetized to the disordered phase of the model. 

This can also be understood by looking at the effective potential experienced by vortices. In the regime where long-range interactions are irrelevant, the effective partition function can be solved exactly, ignoring the long-range exponential dampening. We obtain:
\begin{equation}
    Z\varpropto e^{(4-2\pi \beta_{\rm{eff}})\ln{(L)}}=e^{-V_{\rm{eff}}(L)}.
\end{equation}
Therefore, the effect of long-range interactions can be understood as to renormalize the effective potential experienced by vortices with respect to each other, as portrayed in Figure \ref{fig:figure1}.

\paragraph{Renormalization Group analysis.}\label{sec:rganalysis}
To verify the results obtained in the previous section using an LP argument, we next analyze the model using the perturbative renormalization group. 
% We will work perturbatively in both $g$ and $y$, up to their lowest non vanishing orders. 
The RG equations of the long-range XY model have been calculated to linear order in both $g$ and $y^2$ \cite{giachettidefenu2021}, neglecting terms of the order $gy^2$. Our LP results motivate us to include this additional term, which is responsible for the shift of the critical temperature and modification of the phase diagram.
% We explicitly incorporate interactions proportional to $gy^2$, in order to verify the results obtained by LP arguments. 
We employ the method of the renormalization of the spin wave stiffness \cite{Chaikin_Lubensky_1995}. 
The RG equations of $\beta$, $y$, and $g$ to the relevant orders are
\begin{equation}\label{eq:rgflow_rgequations}
    \begin{split}
        \frac{\partial \beta}{\partial l}=&-\pi^3\beta^2y^2+\pi \beta g\\
        \frac{\partial y^2}{\partial l}=&\left[4-2\pi\beta\left(1-\frac{\pi g}{2-\sigma-\eta}\right)\right]y^2\\
        \frac{\partial g}{\partial l}=&\left[2-\sigma-\eta+\frac{2\pi^2y^2}{4-2\pi \beta}-\pi^3\beta y^2\right]g.
    \end{split}
\end{equation}
Under the full RG flow shown in the End Matter, an additional coupling, $\gamma$, is generated. We neglect the flow of this higher order coupling by setting $\gamma=0$.
% In order to have consistent RG equations, we need to introduce an additional coupling, $\gamma$
% , whose flow we neglect by setting it to 0.
% . At this point, we do not have a physical interpretation of $\gamma$, so we neglect it by setting $\gamma=0$. 
% The complete equations including $\gamma$ can be found in the End Matter. 
%This is an approximation, whose validity and consequences must be resolved by future work. 
% A complete analysis of the equations is quite involved and out of the scope of this work. 

In the quasi-long-range ordered phase we have $y^2=0$. Long-range interactions become relevant for
\begin{equation}
    \sigma<2-\eta,
\end{equation}
which corresponds to a magnetized phase. This can be seen in the second panel of Figure \ref{fig:rgflow}. Let us assume $\sigma>2-\eta$ in the following, so that long-range interactions are irrelevant. To analyze the behavior of vortices, we assume that the physical model has a finite but small value $g>0$. From the RG equation of $y^2$, we see that vortices become relevant for
\begin{equation}
    \beta_{\text{eff}}<\frac{2}{\pi},
\end{equation}
which is the same condition we obtained from the LP argument. Originally, we would recover the long-range XY model with $g=1$, but \textit{a priori} it is not clear whether this is the correct value after the renormalization procedure. The resulting RG flow can be seen in the first panel of Fig.~\ref{fig:rgflow}. We find that long-range interactions enhance the stability of vortices, shifting the BKT transition temperature towards larger temperatures. The resulting phase diagram is the same as obtained by LP arguments, which can be seen in Fig.~\ref{fig:figure1}.

\paragraph{Discussion.}\label{sec:conclusion}
In this Letter, we have found that long-range interactions stabilize the BKT transition, such that even though a magnetized phase emerges, there is no direct transition from the magnetized to the disordered phase of the model. A summary of the phase diagram is shown in Fig.~\ref{fig:figure1}(c). 
To arrive at our results, we have used two complementary methods: First, we have developed an LP argument for a single vortex-antivortex pair. This gives an intuitive understanding of the BKT mechanism in the long-range XY model and describes how a $\betaeff$ emerges in the model. From this, we were able to infer the phase diagram of the model, which we further verified using higher-order RG calculations.
% We built on a recent field theoretic analysis of the model \cite{giachettidefenu2021}, which we crucially expanded to include coupling terms of vortices and long-range interactions. 
% The results from this method agree quantitatively with those obtained using the LP argument.

The phase diagram we obtain in this Letter deviates from previous results on the long-range XY model, specifically, it predicts that the BKT transition persists for arbitrary values of $\sigma$. This is different to the phase diagram obtained in \cite{giachettidefenu2021}, where the BKT transition only exists for $\sigma>7/4$. We explain this effect as being due to the coupling between vortices and long-range interactions.
The fact that the topological excitations can become renormalized by spin waves is a surprising result, which provides a motivation to extend the method to other systems in the future, such as the nematic XY model with an additional term in the Hamiltonian given by $\cos{(2\phi_i-2\phi_j)}$ \cite{Serna_2017, shilamacraftfendley2011} and skyrmion magnets~\cite{garst2017collective}.

%Further, there have been recent numerical studies of the long-range XY model . Using Monte Carlo methods, the authors find that the BKT transition only exists for $\sigma>2$; for $\sigma<2$, they find a second-order phase transition. This is in disagreement with both \cite{giachettidefenu2021} and our results, and to the best of our knowledge, we do not know the source of this disagreement. Using field theoretic methods, long-range interactions are often analyzed perturbatively in the long-range coupling $g$ \cite{giachettidefenu2021, Giachetti2023}. This treatment is verified in the RG procedure, arguing that the long-range coupling is irrelevant in the regime where the BKT transition appears. To resolve the discrepancies between numerical and field theoretic results, it would be interesting to analyze the validity of those approaches, which would hopefully lead to a broader understanding of the long-range XY model and long-range interactions in general. \\

It is surprising that recent works find contradictory results for the phase diagram of such a canonical model of a topological phase transitions. Monte Carlo studies of the long-range XY model \cite{Xiao_2025, yao2025nonclassicalregimetwodimensionallongrange} find that the BKT transition only exists for $\sigma>2$. For $\sigma<2$, they find a second-order phase transition, in disagreement with both Ref.~\cite{giachettidefenu2021} and our results. There are multiple possible explanations for this discrepancy: In their numerical works, the authors normalize the coupling strength of the model as a function of the long-range exponent $\sigma$ to remove finite size corrections, stating that they cannot make quantitative statements about the exact position of the transition. On the contrary, using LP arguments, we cannot infer the universality of the transition, but only the position. Additionally, our results are of perturbative nature in $g$ --- although this approach is validated by the irrelevance of the coupling in the RG, it could be possible that a new critical point emerges at a finite $g$.
Hence, further work is needed to understand the connection between the model with $g=1$ and our perturbative results.
Confirmation could be achieved by studying the long-range XY model with small $g$ numerically, to observe a possible critical change in behavior. Further, a full analysis of the RG equations presented in this letter might shed further light on the behavior of the long-range XY model. 
%
% To resolve the discrepancies between numerical and field theoretic results, further investigation is necessary. 
%In twodimensionalxyferromagnetinduced, yao2025nonclassicalregimetwodimensionallongrange}, the authors renormalize the coupling of the model for different values of $\sigma$, so that they cannot make quantitative statements about the exact transition temperature of the BKT transition. It would be interesting to resolve this issue, to be able to make quantitative statements about the BKT transition temperature and compare our results with numerical calculations. \\
%In our work, we advance the use of LP arguments to determine the behavior of vortices. This method works well for the XY model and similar models, and it would be interesting to benchmark this method further in different models, because it often gives a more intuitive and simple understanding of the behavior of excitations in the model, and it can be used complementarily to known methods, such as RG calculations.\\

While our work is interesting for fundamental reasons, there have also been recent experiments on Rydberg atoms which have been able to implement long-range XY-like interactions \cite{Chen2023rydberg}. There, our predictions could be directly tested in experiment. %There are further experimental platforms, where long-range interactions can be implemented, as for example in Atomic systems \cite{PhysRevLett.108.215301, PhysRevLett.108.210401, Lewis_2023}, Molecular systems \cite{Yan2013}, and Rydberg atoms \cite{RevModPhys.82.2313, Schauß2012, Chen2023rydberg}.
Further, the significance of long-range interactions in the preparation of exotic quantum phases and for quantum computation protocols has recently been demonstrated \cite{Lewis_2023, Chen2023rydberg}, which provides a complementary motivation for a better understanding of the interplay between long-range interactions and topological excitations.

\acknowledgments{\emph{Acknowledgments.}} We would like to thank M. Knap, J. Habel, and P. Rao for insightful discussions.
J.K. acknowledges support from the Deutsche Forschungsgemeinschaft (DFG, German Research Foundation) under Germany’s Excellence Strategy (EXC–2111–390814868 and ct.qmat EXC-2147-390858490), and DFG Grants No. KN1254/1-2, KN1254/2-1 TRR 360 -- 492547816 and SFB 1143 (project-id 247310070), as well as the Munich Quantum Valley, which is supported by the Bavarian state government with funds from the Hightech Agenda Bayern Plus. J.K. further acknowledges support from the Imperial-TUM flagship partnership.

\bibliography{main}

% \newpage
\appendix
\section{End Matter}
\subsection{Renormalized BKT transition temperature}
The renormalized BKT transition temperature reads
\begin{equation}
    \beta_{\rm{eff}}=\beta\left(1-\frac{\pi g}{2-\sigma-\eta}\right).
\end{equation}
The BKT transition appears for the critical temperature $\beta_{\rm{eff}, c}=2/\pi$. This condition can be solved analytically, and as a result we obtain for the critical temperature $T_{{\mathrm{BKT}}}=1/\beta_c$:
\begin{equation}
    T_{\mathrm{BKT}}= \frac{\pi}{4}\left(9-4\sigma +\sqrt{49+16\pi g-56\sigma +16\sigma^2}\right).
\end{equation}
\subsection{Critical temperature of long-range interactions}
Long-range interactions become relevant when
\begin{equation}
    \sigma<2-\eta.
\end{equation}
Recalling that $\eta=T/2\pi$, we can solve for the temperature, and we obtain
\begin{equation}
    T_{\mathrm{c, lr}}=2\pi(2-\sigma).
\end{equation}
\subsection{Remaining Integral in the LP argument}\label{sec:app_lp_integral}
After integrating out spin waves, we are left with evaluating the following integral over a vortex-antivortex configuration
\begin{equation}\label{eq:endmatter_integral}
    I\left(r\right)=\!\!\!\int\displaylimits_{1<|\bm{x}-\bm{y}|}\!\!\!\!\!\frac{\dd[2]\bm{x}\dd[2]\bm{y}}{|\bm{x}-\bm{y}|^{2+\sigma+\eta}}\left[1\!-\!\cos{\big(\theta(x)\!-\!\theta(y)\big)}\right].
\end{equation}
Here, $\theta(\bm{x})$ describes the field of a vortex-antivortex configuration with separation $r$ between the vortex and antivortex. All distances are measured in units of the lattice spacing $a$. To evaluate the integral, we split it into two regions:
\begin{equation}\label{eq:endmatter_integralsplit}
    \begin{split}
        I\left(r\right)=&\!\!\!\int\displaylimits_{1<|\bm{x}-\bm{y}|<r}  \!\!\!\!\!\frac{\dd[2]\bm{x}\dd[2]\bm{y}}{|\bm{x}-\bm{y}|^{2+\sigma+\eta}}\left[1\!-\!\cos{\big(\theta(\bm{x})\!-\!\theta(\bm{y})\big)}\right]\\
        +&\!\!\!\int\displaylimits_{r<|\bm{x}-\bm{y}|} \!\!\!\!\! \frac{\dd[2]\bm{x}\dd[2]\bm{y}}{|\bm{x}-\bm{y}|^{2+\sigma+\eta}}\left[1\!-\!\cos{\big(\theta(\bm{x})\!-\!\theta(\bm{y})\big)}\right].
    \end{split}
\end{equation}
We are going to evaluate both summands separately. We start with the first summand. Here, we have restricted $1<|\bm{x}-\bm{y}|<r$. In this regime, the field $\theta$ is smooth enough to expand the cosine to second order. We obtain
\begin{equation}
\begin{split}
    &\int\displaylimits_{1<|\bm{x}-\bm{y}|<r}   \!\!\!\!\!\frac{\dd[2]\bm{x}\dd[2]\bm{y}}{|\bm{x}-\bm{y}|^{2+\sigma+\eta}}\left[1-\cos{\big(\theta(\bm{x})-\theta(\bm{y})\big)}\right]\\
    =&\frac{1}{2}\!\!\!\!\!\int\displaylimits_{1<|\bm{x}-\bm{y}|<r} \!\!\!\!\!\frac{\dd[2]\bm{x}\dd[2]\bm{y}}{|\bm{x}-\bm{y}|^{2+\sigma+\eta}}\left((\bm{x}-\bm{y})\cdot \bm{\nabla}\theta(\bm{x})\right)^2\\
    =&\frac{1}{2}\int \dd[2]\bm{x}\left(\bm{\nabla}\theta(\bm{x})\right)^2\!\!\!\int \displaylimits_{1<|\bm{q}|<r} \!\!\!\frac{\dd[2]\bm{q}}{|\bm{q}|^{\sigma+\eta}}\cos{(\alpha)}^2 \\
    =&\frac{\pi}{2(2-\sigma-\eta)}\left(r^{2-\sigma-\eta}-1\right)\int \dd[2]\bm{x} \left(\nabla\theta(\bm{x})\right)^2 \\
    =&\frac{2\pi^2}{2-\sigma-\eta}\left(r^{2-\sigma-\eta}-1\right)\ln{\left(r\right)}.
\end{split}
\end{equation}
Here, we used that:
\begin{equation}
    \int d^2x \left(\nabla\theta(x)\right)^2=4\pi\ln{\left( r\right)}.
\end{equation}
The expansion of the cosine for $|\bm{x}-\bm{y}|<r$ is an apparent approximation. Nevertheless, the argument as to why this is allowed is similar to the short-range XY model: On distances smaller than the scale $r$, the vortex field configuration in the system is smooth, so that the cosine can be approximated by the quadratic expansion. Also, note that the exact length $r$ where we split the integral does not matter: we could have chosen $1<|x-y|<r/m$ with $m>1$, and the calculation is similar. This would reduce the error of this approximation, at the cost of introducing the new length scale $r/m$.

Next, we consider the second summand of Eq.~\eqref{eq:endmatter_integralsplit}. Here, we can simply rescale $\bm{x}$ and $\bm{y}$ by $r$ to obtain
\begin{equation}
    \int\displaylimits_{r<|\bm{x}-\bm{y}|<L}  \!\!\!\!\!\!\!\!\frac{\dd[2]\bm{x}\dd[2]\bm{y}}{|\bm{x}-\bm{y}|^{2+\sigma+\eta}}\left[1\!-\!\cos{\big(\theta(\bm{x})\!-\!\theta(\bm{y})\big)}\right]=k\cdot r^{2-\sigma-\eta}.
\end{equation}
Here, $k$ is a positive, $r$-independent constant. Combining these results, we have
\begin{equation}\label{eq:lrxy_lp_theintegralfullsolution}
\begin{split}
    I(r)=&J\left(r\right)\cdot r^{2-\sigma-\eta}-\frac{2\pi^2}{2-\sigma-\eta}\cdot \ln{\left(r\right)}.
\end{split}
\end{equation}
Here, we have defined
\begin{equation}\label{eq:lrxy_lp_thejfunction}
    J\left(r\right)=\frac{2\pi^2}{2-\sigma-\eta}\ln{\left(r\right)}+k,
\end{equation}
which only has a weak $r$-dependence proportional to $\ln{\left(r\right)}$. Since $J(r)$ is multiplied with a polynomial in $r$, this $r$ dependence does not determine the asymptotic behavior of $I(r)$ and can be neglected.

\subsection{RG flow equations}
We derive the RG equations of the long-range XY model by utilizing the method of spin wave renormalization \cite{Chaikin_Lubensky_1995}. The advantage of this method is that it does not rely on duality mappings, which we know are not applicable in the long-range XY model \cite{Giachetti2023Villain}. To start, we impose an external gradient on the field $\phi(\bm{x})$ as follows:
\begin{equation}
    \phi(\bm{x})=\Tilde{\phi}(\bm{x})+\bm{w}\cdot \bm{x}.
\end{equation}
We define the macroscopic renormalized temperature as the curvature of the free energy with respect to $\bm{w}$ as
\begin{equation}\label{eq:endmatter_rencon}
    \frac{1}{2}L^2\beta_R \bm{w}^2=\beta F(\bm{w})-\beta F(\bm{0}).
\end{equation}
The free energy is defined as
\begin{equation}\label{eq:endmatter_freeenergy}
    \beta F(\bm{w})=-\ln{\left\{\int D[\varphi]\sum\displaylimits_{n_{\bm{r}}}\exp\left[-\beta H(\bm{w})\right] \right\}}.
\end{equation}
Here, $H(\bm{w})$ is the Hamiltonian of the long-range XY model from Eq.~\eqref{eq:hamil_writtenas_srandlr} including the field gradient $\bm{w}$. We define the spin wave and vortex averages as
\begin{align}
    \langle O[\varphi]\rangle_{\mathrm{sw}}=&\frac{1}{Z_{\mathrm{sw}}}\int D[\varphi]e^{-\frac{\beta}{2}\int \dd \bm{x}(\bm{\nabla}\varphi(\bm{x})^2}O[\varphi]\\
    \langle O[n_{\bm{r}}]\rangle_{v}=&\frac{1}{Z_{v}}\sum \displaylimits_{n_{\bm{r}}}y^{N}e^{-\beta\pi\sum_{\bm{x}, \bm{y}}n_{\bm{x}}n_{\bm{y}}G(\bm{x}-\bm{y})}O[n_{\bm{r}}],
\end{align}
where $N=\sum_{\bm{r}}n_{\bm{r}}^2$ is the sum over squared charges of vortices. $Z_{\mathrm{sw}}$ and $Z_v$ are defined by $\langle 1\rangle_{\mathrm{sw}}=\langle 1\rangle_v=1$.

Using this, the free energy Eq.~\eqref{eq:endmatter_freeenergy} becomes
\begin{widetext}
    \begin{equation}
    \beta F(\bm{w})=\ln{\left\{\left\langle e^{-\beta \int \dd[2]\bm{r} \hspace{1mm}\bm{w}\cdot \bm{\nabla}\theta(\bm{x})}e^{-\beta g\int\frac{\dd[2]\bm{x}\dd[2]\bm{y}}{|\bm{x}-\bm{y}|^{2+\sigma}}\left[1-\cos{(\varphi(\bm{x})-\varphi(\bm{y})+\theta(\bm{x})-\theta(\bm{y})+\bm{w}\cdot (\bm{x}-\bm{y}))}\right]}\right\rangle_{v, sw}Z_{\mathrm{sw}}Z_v\right\}}.
\end{equation}
\end{widetext}
We recall that 
\begin{equation}
    \theta(\bm{x})=\sum_{\bm{r}}n_{\bm{r}}\cdot \arg((x_1-r_1)+i(x_2-r_2)),
\end{equation}
where $\arg(c)$ is the complex argument function which gives the polar phase of the complex number $c$. Here and in the following we measure distances in units of the lattice spacing $a$, effectively setting $a=1$.
To make analytical progress, we need to expand the free energy to linear order in $g$. Further, we are only interested in $F(\bm{w})$ to second order in $\bm{w}$, so we can also expand in $\bm{w}$. Due to the isotropy of the system, the first order in $\bm{w}$ vanishes and the second order only depends on $\bm{w}^2$.
When only keeping the relevant terms, we obtain the following equation for $\beta_R$:
\begin{widetext}
    \begin{equation}\label{eq:endmatter_betarlong}
    \begin{split}
        \beta_R=&\beta -\frac{\beta^2}{2L^2}\int \dd[2]\bm{r}\dd[2]\bm{r}'\big\langle \bm{\nabla}\theta(\bm{r})\cdot \bm{\nabla}\theta(\bm{r}')\big\rangle_v\left[1-\beta g\int \frac{\dd[2]\bm{x}\dd[2]\bm{y}}{|\bm{x}-\bm{y}|^{2+\sigma}}\langle \cos{(\varphi(\bm{x})-\varphi(\bm{y})}\rangle_{\mathrm{sw}} \right]\\
        &+\frac{\beta g}{2L^2}\int \frac{\dd[2]\bm{x}\dd[2]\bm{y}}{|\bm{x}-\bm{y}|^{\sigma}}\langle \cos{(\varphi(\bm{x})-\varphi(\bm{y})}\rangle_{\mathrm{sw}}\left\langle\cos{(\theta(\bm{x})-\theta(\bm{y}))}\left[1-\frac{\beta^2}{|\bm{x}-\bm{y}|^2}\int \dd[2]\bm{r}\dd[2]\bm{r}'\bm{\nabla}\theta(\bm{r})\cdot \bm{\nabla}\theta(\bm{r}')\right]\right\rangle_v\\
        &-\frac{\beta^2 g}{L^2}\int \frac{\dd[2]\bm{x}\dd[2]\bm{y}}{|\bm{x}-\bm{y}|^{2+\sigma}}\langle \cos{(\varphi(\bm{x})-\varphi(\bm{y})}\rangle_{\mathrm{sw}}\int\dd[2]\bm{r}(\bm{x}-\bm{y})\cdot\langle\bm{\nabla}\theta(\bm{r}) \sin{(\theta(\bm{x})-\theta(\bm{y}))}\rangle_v
    \end{split}
    \end{equation}
\end{widetext}
We recall that $\langle\cos{(\varphi(\bm{x})-\varphi(\bm{y}))}\rangle_{\mathrm{sw}}=1/|\bm{x}-\bm{y}|^{\eta}$. Further, the expectation value over vortices can be calculated perturbatively in $y$. We define $\langle O\rangle_N$ as the expectation value of the operator $O$ with respect to vortices including all terms to order $y^N$, and correspondingly we define $Z_N$. Then, for operators $O$ that vanish when no vortices are present, we have that
\begin{equation}
    \langle O\rangle_2=y^2\int\frac{\dd[2]\bm{r}_1\dd[2]\bm{r}_2}{|\bm{r}_1-\bm{r}_2|^{2\pi\beta}}O(\bm{r}_1, \bm{r}_2).
\end{equation}
Here, $\bm{r}_1$ is the location of the vortex and $\bm{r}_2$ the location of the antivortex. Using this, we can transform Eq.~\eqref{eq:endmatter_betarlong} into a more tractable form. While doing so, integrals of the following form appear:
\begin{equation}
    \int \frac{\dd[2]\bm{x}\dd[2]\bm{y}}{|\bm{x}-\bm{y}|^{u}}\int \frac{\dd[2]\bm{r}_1\dd[2]\bm{r}_2}{|\bm{r}_1-\bm{r}_2|^{u}}(...).
\end{equation}
To solve those integrals, we split the integration domain into the regions $|\bm{x}-\bm{y}|<|\bm{r}_1-\bm{r}_2|$ and $|\bm{x}-\bm{y}|>|\bm{r}_1-\bm{r}_2|$. To solve the integrals in both regions, we note that the vortex phase $\theta$ fulfills a neat duality condition. Denoting by $\theta(\bm{x}, \bm{r}_1)$ the field at position $\bm{x}$ caused by a vortex at position $\bm{r}_1$, we have that $\theta(\bm{x}, \bm{r}_1)=\theta(\bm{r}_1, \bm{x})+\pi$. Therefore, the difference of vortex phases which appear in Eq.~\eqref{eq:endmatter_betarlong} have the following property:
% \begin{widetext}
%     \begin{equation}
%         \theta(\bm{x})-\theta(\bm{y})=\theta(\bm{x}, \bm{r}_1)-\theta(\bm{x}, \bm{r}_2)-\theta(\bm{y}, \bm{r}_1)+\theta(\bm{y}, \bm{r}_2)=(\theta(\bm{r}_1, \bm{x})-\theta(\bm{r}_1, \bm{y}))-(\theta(\bm{r}_2, \bm{x})-\theta(\bm{r}_2, \bm{y})).
%     \end{equation}
% \end{widetext}
\begin{multline}
    \theta(\bm{x})-\theta(\bm{y})
    =\theta(\bm{x}, \bm{r}_1)-\theta(\bm{x}, \bm{r}_2)-\theta(\bm{y}, \bm{r}_1)+\theta(\bm{y}, \bm{r}_2) \\
    =(\theta(\bm{r}_1, \bm{x})-\theta(\bm{r}_1, \bm{y}))-(\theta(\bm{r}_2, \bm{x})-\theta(\bm{r}_2, \bm{y})).
\end{multline}
Using this, the appearing integrals in Eq.~\eqref{eq:endmatter_betarlong} can be solved analogously to the integral in Eq.~\eqref{eq:endmatter_integral}. Further, we use that 
\begin{equation}
    1+\alpha \ln{(r)}=r^{\alpha}+\mathcal{O}(\alpha^2).
\end{equation}
Then, introducing back the lattice spacing $a$, Eq.~\eqref{eq:endmatter_betarlong} becomes
\begin{widetext}
    \begin{equation}\label{eq:endmatteer_betafinal}
    \begin{split}
        \beta_R=&\beta-\pi^3\beta^2y^2e^{-\frac{2\pi g}{2-\sigma-\eta}}\int_a^{\infty}\frac{\dd{r}}{a}\left(\frac{r}{a}\right)^{3-2\pi\beta\left(1-\frac{\pi g}{2-\sigma-\eta}\right)}+\pi \beta ge^{\frac{2\pi^3\beta y^2}{4-2\pi\beta}}\int \displaylimits_{a}^{\infty}\frac{\dd{r}}{a} \left(\frac{r}{a}\right)^{1-\sigma-\eta+\frac{2\pi^2y^2}{4-2\pi\beta}}\\
        &+\beta g y^2\pi^3\int \displaylimits_a^{\infty}\frac{\dd{r}}{a}\left(\frac{r}{a}\right)^{5-\sigma-\eta-2\pi\beta}\left(\frac{-2\beta \pi}{2-\sigma-\eta}+\frac{-2\beta\pi}{4-2\pi\beta}+\ln{\left(\frac{r}{a}\right)}\left(\frac{2\beta^2\pi^2}{2-\sigma-\eta}-\frac{2}{4-2\pi\beta}\right)\right).
    \end{split}
    \end{equation}
\end{widetext}
All integrals in Eq.~\eqref{eq:endmatteer_betafinal} come with the lattice spacing as their lower cutoff. We renormalize the theory by introducing a second scale, which is slightly larger than the lattice spacing $a\rightarrow a'>a$ and introducing new variables $\beta', g', y'$, while keeping the macroscopic $\beta_R$ fixed.  In the case of an infinitesimal change $a'=ae^{dl}$ with $dl\ll1$ this leads to the differential renormalization group equations:
\begin{equation}
    \begin{split}
        \frac{\partial\beta}{\partial l}\!=&\!-\pi^3\beta^2y^2\!+\!\pi \beta g\!-\!\gamma \beta g y^2\pi^3\!\left(\!\frac{2\beta \pi}{2-\sigma-\eta}\!+\!\frac{2\beta\pi}{4-2\pi\beta}\!\right)\\
        \frac{\partial y^2}{\partial l}\!=&\!\left(4-2\pi\beta\left(1-\frac{\pi g}{2-\sigma-\eta}\right)\right)y^2\\
        \frac{\partial g}{\partial l}\!=&\!\left(2-\sigma-\eta+\frac{2\pi^2y^2}{4-2\pi \beta}-\pi^3\beta y^2\right)g\\
        \frac{\partial \gamma}{\partial l}\!=&\!-\frac{\beta^2\pi^2(4-2\pi\beta)-(2-\sigma-\eta)}{\beta \pi(6-2\pi\beta-\sigma-\eta)}.
    \end{split}
\end{equation}
The additional coupling $\gamma$ is generated under the RG flow. The appearance can be understood by the fact that out of the RG procedure we obtain four equations (one for $\beta$, $y^2$, $g$ and $gy^2$), but we only have three couplings to renormalize, so the system is overdetermined. We do not have a physical interpretation of $\gamma$, and henceforth set $\gamma=0$ in the analysis of the equations. The analysis of the full RG equations including $\gamma$ is left open for future work.

\end{document}